\documentclass[letter,11pt]{article}
 \usepackage{jheppub}
 
 \usepackage{graphicx}
 \usepackage{hyperref}
 \usepackage[utf8]{inputenc}
 \usepackage{amssymb}
\usepackage{multirow}
\usepackage{soul}

\usepackage{indentfirst}
\usepackage{graphicx}
\usepackage{dcolumn}
\usepackage{bm}
\usepackage{braket}
\usepackage{natbib}

\usepackage{hyperref}
\usepackage{booktabs}
\usepackage{mathrsfs}
\usepackage{epsfig}
\usepackage{graphicx}
\usepackage{subfigure}
\usepackage{dcolumn}
\usepackage{bm}
\usepackage{amsmath}
\usepackage{slashed}
\usepackage{multirow}
\usepackage{color}
\usepackage{dcolumn}
\usepackage{bm}
\usepackage{color}

\allowdisplaybreaks

 \newcommand{\lsim}{{\;\raise0.3ex\hbox{$<$\kern-0.75em\raise-1.1ex\hbox{$\sim$}}\;}}
\newcommand{\gsim}{{\;\raise0.3ex\hbox{$>$\kern-0.75em\raise-1.1ex\hbox{$\sim$}}\;}}
\newcommand{\beq}{\begin{equation}}
\newcommand{\eeq}{\end{equation}}
\newcommand{\bea}{\begin{eqnarray}}
\newcommand{\eea}{\end{eqnarray}}

\def\baa{\begin{array}}
\def\eaa{\end{array}}

\mathchardef\minus="002D

\preprint{ }

\title{ Superhorizon fluctuations and the cosmic dipole problem}

\author{Ge Chen$^{1}$, }

\author{Chengcheng Han$^{1,2}$,}
\emailAdd{hanchch@mail.sysu.edu.cn}

\author{Linwei Qiu$^{1}$}

\vspace{0.5cm}

\affiliation{$^1$School of Physics, Sun Yat-Sen University, Guangzhou 510275, P. R. China}
\affiliation{$^2$Asia Pacific Center for Theoretical Physics, Pohang 37673, Korea}

\vspace{0.5cm}

\abstract{ Recent observations have identified a significant 4.9$\sigma$ tension between the cosmic dipole inferred from galaxy number counts and that derived from the Cosmic Microwave Background (CMB), suggesting a potential deviation from the cosmological principle. This work investigates whether superhorizon isocurvature perturbations in cold dark matter (CDM) can account for this discrepancy. We demonstrate that, unlike adiabatic modes which cancel at leading order, superhorizon isocurvature modes can generate an intrinsic CMB dipole without significantly affecting galaxy number counts, thereby explaining the observed mismatch. We explore both single-mode and continuous-spectrum cases, focusing on two concrete models: a nearly scale-invariant power-law spectrum with a UV cutoff and axion-induced isocurvature perturbations. For the axion scenario, we show that if the radial mode evolves during inflation, the resulting perturbations can match the required amplitude while evading current CMB constraints. Our analysis constrains the self-coupling of associated potential for the axion to the range $10^{-9} < \lambda < 4 \times 10^{-9}$. These findings offer a viable solution to the dipole tension and may serve as indirect evidence for axion dark matter.}

\makeatletter
\def\@fpheader{\relax}
\makeatother

\date{\today}

\begin{document} 
\maketitle
\flushbottom
\newpage

\section{\label{sec:intro}INTRODUCTION}
The cosmological principle asserts that the universe is homogeneous and isotropic on large scales, which is a fundamental assumption underpinning many key results in modern cosmology. Ellis and Baldwin~\cite{ellis_expected_1984} note that the cosmic rest frame~(CRF) of galaxies and dark matter should coincide with that of the Cosmic Microwave Background~(CMB) if the cosmological principle holds, which is a powerful consistency test of the Friedman-Lema{\^\i}tre-Robertson-Walker~(FLRW) models. If the observed CMB dipole anisotropy is indeed kinematic due to our local peculiar motion to the CRF, then there must be a corresponding kinematic dipole imprinted in the galaxy number counts on the sky~\cite{ellis_expected_1984, condon_nrao_1998, maartens_kinematic_2018, nadolny_new_2021, gibelyou_dipoles_2012}.

However, recent studies have found a significant discrepancy between the cosmic dipole inferred from galaxy number counts and that derived from the CMB, suggesting that the two CRFs do not align~\cite{mauch_sumss_2003, itoh_dipole_2010, singal_large_2011, gibelyou_dipoles_2012, rubart_cosmic_2013, tiwari_revisiting_2016, singal_large_2019, siewert_cosmic_2021, secrest_test_2021, secrest_challenge_2022}. 
Notably, distant radio galaxies and quasars exhibit a dipole amplitude 2-3 times larger than predicted by the kinematic interpretation of the CMB dipole, with the highest reported tension now approximately $4.9\sigma$~\cite{secrest_test_2021,secrest_challenge_2022, Secrest:2025wyu}. Such results raise doubts regarding the validity of the cosmological principle.

It is conceivable that matter may exhibit a relative velocity to the CMB if a significant superhorizon cold dark matter~(CDM) isocurvature perturbation exists~\cite{erickcek_scale-dependent_2009, domenech_galaxy_2022}. Turner~\cite{turner_tilted_1992} proposed that such a superhorizon isocurvature fluctuation could introduce an intrinsic dipole modulation in the CMB, referring to this case as a “tilted universe” to suggest that galaxies would appear to undergo a global motion relative to the CMB rest frame. Subsequent studies have explored the effects of superhorizon fluctuations on the CMB in detail~\cite{erickcek_superhorizon_2008, domenech_galaxy_2022}, establishing that only isocurvature modes can generate a leading-order CMB dipole.

To address the cosmic dipole problem through a significant superhorizon isocurvature perturbation, a concrete model capable of generating such isocurvature modes is required. In \cite{Domenech:2022mvt}  a single mode of isocurvature perturbation is studied to resolve the cosmic dipole problem. In this paper, we consider the case of isocurvature as a continuum spectrum. Particularly, we consider two models: one is the isocurvature power spectrum follows a nearly scale-invariant power law with,

\begin{equation}
   \mathcal{P}_S(k) = A_I (k/k_\text{max})^{n-1}\theta(k_\text{max}-k)~,
\end{equation}
where $\theta$ is the Heaviside step function. To be consistent with the isocurvature observation at the CMB, for the first model we require the power spectrum has the UV cut off at a particular large scale $k_\text{max}$. The other case is that the isocurvature is generated by the axion dark matter proposed in \cite{han_qcd_2023}. 
Axion is the hypothetical particles arising from the spontaneous breaking of the Peccei-Quinn (PQ) symmetry, a mechanism originally introduced to resolve the strong CP problem in quantum chromodynamics (QCD)~\cite{peccei_cp_1977, peccei_constraints_1977}. If the PQ symmetry is spontaneously broken during the inflationary epoch, axion dark matter can generate isocurvature perturbations with an amplitude determined by the Hubble scale, $H/(\pi f_a \theta_0)$~\cite{steinhardt_saving_1983, preskill_cosmology_1983, dine_not-so-harmless_1983, abbott_cosmological_1983, seckel_isothermal_1985, lyth_limit_1990, turner_inflationary_1991, linde_axions_1991, kim_axions_2009, wantz_axion_2010, marsh_axion_2016, han_qcd_2023, chen_new_2024}. These perturbations are distinct from and uncorrelated with primordial perturbations. Due to the absence of detected isocurvature modes in current CMB observations, a stringent constraint exists on the axion dark matter fraction and the amplitude of axion isocurvature perturbations~\cite{collaboration_planck_2021}. Consequently, if the axion dark matter generates large isocurvature perturbations on superhorizon scales during inflation, which must decay to sufficiently small levels at the recombination scale, $k_\text{dec}$~\cite{han_qcd_2023}.

In this paper, we provide a brief overview of galaxy number counts, CMB temperature fluctuations, and the cosmic dipole tension in Section~\ref{sec:review}. In Section~\ref{sec:signle mode} we examine these fluctuations in the presence of a single superhorizon mode, considering both adiabatic and CDM isocurvature modes. In Section~\ref{sec:continuous spectrum}, we extend the analysis to a continuous spectrum of fluctuations for a more realistic scenario, where we evaluate two models: one power-law spectrum and the other based on the axion isocurvature model.

\section{\label{sec:review}THE COSMIC DIPOLE PROBLEM: A REVIEW}
Due to our peculiar velocity, a dipole anisotropy appears in the observed Cosmic Microwave Background (CMB). From measurements of this CMB dipole,
\begin{align}
D_1^{\rm CMB} = (1.23357 \pm 0.00036) \times 10^{-3}~, 
\end{align}
our peculiar velocity is inferred to be
\begin{align}
|v_\odot^i| = 369.82 \pm 0.11~ {\rm km/s}~.
\end{align}
This velocity also induces a dipole in the galaxy number counts, given by
\begin{align}
D_{\text{kin}}^\text{N}(\mathbf{n}) = [2+x(1+\alpha)] v_\odot^i n_i~,
\end{align}
where the parameter $x$ characterizes the cumulative number of sources above a flux threshold, following the relation $N(>S) \propto S^{-x}$, and $\alpha$ is the spectral index of the galaxy sources, assuming a power-law luminosity spectrum $L(\nu) \propto \nu^{-\alpha}$. 
For the CatWISE quasar catalogue~\cite{secrest_test_2021, secrest_challenge_2022}, the measured average values are $\braket{x} \simeq 1.7$ and $\braket{\alpha} \simeq 1.26$. Based on these parameters, if both the CMB dipole and the galaxy number count dipole are entirely due to our peculiar velocity $v_\odot$, the galaxy dipole amplitude is expected to be approximately four times larger than that of the CMB dipole. However, the observed galaxy dipole is about twice as large as this prediction, 
\begin{align}
D_{\text{kin}}^\text{N}(\mathbf{n}) =(15.54 \pm 1.7)\times 10^{-3}~,
\end{align}
although it points in a similar direction.
This suggests a peculiar velocity of
\begin{align}
v_\odot^i = 797 \pm 87~ {\rm km/s}~,
\end{align}
which deviates by about $5\sigma$ from the velocity inferred from the CMB dipole measurement.

To account for this discrepancy, we consider the possibility that it originates from superhorizon perturbations. Working in the conformal Newtonian gauge, we adopt the perturbed Friedmann–Robertson–Walker (FRW) metric:
\begin{equation}
ds^2 = a^2(\tau)\left[-(1+2\Psi),d\tau^2 + (1-2\Phi),\delta_{ij},dx^i,dx^j\right]~.
\end{equation}
For both the CMB and galaxy number count fluctuations, the total density fluctuation can be decomposed into two components: a kinematic part, arising from the peculiar motion of the observer, and an intrinsic part, generated by the perturbations themselves. Accordingly, the fluctuation can be expressed as
\begin{equation}
\Delta(\mathbf{n},z) = D_\text{kin}(\mathbf{n}) + \Delta_\text{in}(\mathbf{n},z)~,
\end{equation}
where the kinematic component, induced by the observer's motion relative to the cosmic rest frame (CRF), contributes only to the dipole term. The CMB temperature fluctuation is then given by~\cite{erickcek_superhorizon_2008, durrer_cosmic_2020}
\begin{align}
    \Delta^\text{CMB}_\text{in}(\mathbf{n}) &= \frac{1}{4}\delta_r + \Psi + \int_{0}^{r_\text{dec}} dr\ (\Phi' + \Psi') + v_o^i n_i - v^i_s n_i~, \label{equ:in_CMB}\\
    D_{\text{kin}}^\text{CMB}(\mathbf{n}) &= v_\odot^i n_i~, \label{equ:kin_CMB}~,
\end{align}
where $\delta_r$ is the density contrast of radiation, $v_s$ is the velocity potential at source, $v_i =\partial_i v$ and $v_o = v(\tau=0)$.
And the galaxy number count fluctuation is given by~\cite{bonvin_what_2011, challinor_linear_2011, ellis_expected_1984}
\begin{align}
    &\begin{aligned}
        \Delta^\text{N}_\text{in}(\mathbf{n},z) &= \left(2 + \frac{\mathcal{H}'}{\mathcal{H}^2}+\frac{2-5s}{r_s\mathcal{H}} - f_\text{evo} \right) {v}_o^i {n}_i \\ 
        &+ b\delta_{mc} + (3-f_\text{evo})\mathcal{H}v_s + \Psi - (2-5s)\Phi + \frac{1}{\mathcal{H}}[\Phi' + \partial_r({v}_s^i {n}_i)] \\
        &+ \left(\frac{\mathcal{H}'}{\mathcal{H}^2}+\frac{2-5s}{r_s\mathcal{H}}+5s-f_\text{evo}\right)\left(\Psi + {v}_s^i {n}_i + \int_0^{r_s} (\Psi'+\Phi') dr\right) \\
        &+ \frac{2-5s}{2r_s}\int_0^{r_s} \left(2-\frac{r_s-r}{r}\nabla^2_\Omega\right)(\Psi+\Phi) dr~,
    \end{aligned} \label{equ:in_NC}\\
    &D_{\text{kin}}^\text{N}(\mathbf{n}) = [2+x(1+\alpha)] v_\odot^i n_i~,
    \label{equ:kin_NC}
\end{align}
where $\delta_{mc}$ is the comoving density contrast of matter. $b$, $f_\text{evo}$ and $s$ are the galaxy bias, the magnification bias and the evolution bias respectively, which depend on the specifics of the galaxy survey and the instrument.

When focusing on superhorizon perturbations, the dipole induced by adiabatic modes cancels out in both the CMB and galaxy number count dipoles. However, for superhorizon isocurvature modes, an additional dipole arises in the CMB but not in the galaxy number counts~\cite{erickcek_superhorizon_2008, domenech_galaxy_2022}.
This distinction can help explain the observed discrepancy between the CMB dipole and the galaxy number count dipole. Specifically, superhorizon isocurvature modes can contribute a negative intrinsic dipole to the CMB, partially canceling the kinematic dipole caused by our motion. Since these modes do not significantly affect galaxy number counts, the galaxy dipole still accurately reflects our true velocity. As a result, the observed CMB dipole appears smaller than expected, thereby helping to resolve the discrepancy~\cite{domenech_galaxy_2022}.

Accordingly, the CMB dipole power can be written as
\begin{equation}
C_1^{\text{CMB}} = C_{\text{kin},1}^{\text{CMB}} - C_{\text{in},1}^{\text{CMB}}~,
\end{equation}
where the intrinsic dipole component is constrained by the angular power spectrum to lie within the range~\cite{han_qcd_2023}
\begin{equation}
9.0 \times 10^{-7} \lesssim C_{\text{in},1} \lesssim 5.6 \times 10^{-6}~.
\end{equation}
This allows for a naturally suppressed CMB dipole relative to the galaxy number count dipole. It is important to note that, in realistic cosmological models, perturbations typically originate from a continuous spectrum rather than being dominated by a single mode. In this work, we will analyze the effects of isocurvature perturbations in both cases.

\section{\label{sec:signle mode}THE EFFECT OF A SINGLE PLANE WAVE}
For a single superhorizon mode with a given $k$, the metric perturbation can be written as $\Psi(x,\tau) = \Psi_k(\tau) e^{i\mathbf{k}\cdot\mathbf{r}}$.
Applying Rayleigh's plane wave expansion that
\begin{equation}
  e^{ik\cdot r} = \sum_{l=0}^\infty \sum_{m=-l}^l a_{lm} Y_{lm}(\hat{\mathbf{r}}) = 4\pi \sum_{l=0}^\infty \sum_{m=-l}^l i^l j_l(kr) Y^*_{lm}(\hat{\mathbf{k}}) Y_{lm}(\hat{\mathbf{r}})~,
\end{equation}
where $j_l(x)$ is the spherical Bessel function and $Y_{lm}(\hat{\mathbf{r}})$ is the spherical harmonics.
We then have
\begin{equation}
    \begin{aligned}
      a_{lm}^\mathcal{I} &= 4\pi i^l X^\mathcal{I} \mathcal{T}_l^\mathcal{I}(k) Y_{lm}^*(\hat{\mathbf{k}})~, \\
      a_{lm} &= 4\pi i^l \left(\sum_\mathcal{I} X^\mathcal{I} \mathcal{T}_l^\mathcal{I}(k)\right) Y_{lm}^*(\hat{\mathbf{k}})~,
    \end{aligned}
\end{equation}
where $\mathcal{I}$ labels the perturbation modes correspond to different initial condition $\mathcal{I}$ and $\mathcal{T}^\mathcal{I}$ is the corresponding transfer function of $X$ that
$X(\tau,k) = \mathcal{T}(\tau,k)X_\text{ini}$.

Without loss of generality, consider a in $z$-direction plane wave fluctuation, we can expand it in both powers of $(kr_\text{dec} \cos\theta)$ and spherical harmonics.
For $\hat{\mathbf{k}} = \hat{\mathbf{z}}$, all the $Y_{lm}$s for $m \neq 0$ will then vanish and thus
\begin{equation}
    \Delta(\hat{\mathbf{r}},z) = \sum_l \frac{1}{l!} D_l (kr_\text{dec} \cos\theta)^n = \sum_l a_{l0} Y_{l0}(\theta,\varphi)~.
\end{equation}
Comparing the coefficients of each $\cos^l\theta$ we can express $D_l$ in terms of the $a_{l0}$s for the z-direction plane wave.

\subsection{CMB fluctuations}
Defining $\mathcal{T}_{X}^\mathcal{I}(\tau,k)$ as the transfer function of some variable $X$ at conformal time $\tau$ and comoving wave number $k$, we have 
\begin{equation}
  X^\text{ad}(\tau,k) = \mathcal{T}_{X}^\text{ad}(\tau,k) \Phi_\text{ini} \quad \text{or} \quad X^\text{iso}(\tau,k) = \mathcal{T}_{X}^\text{iso}(\tau,k) S_\text{ini}~,
\end{equation}
for adiabatic or isocurvature perturbation. 
Substituting these transfer functions into Eq.~(\ref{equ:in_CMB}) and (\ref{equ:kin_CMB}), we find that the spherical harmonic coefficients, $a_{lm}$, become
\begin{equation}
  a_{lm} = 4\pi i^l \left[F_l^\text{ad}(k) \Phi_\text{dec} + F_l^\text{iso}(k) S_\text{dec}\right] Y_{lm}^*(\hat{\mathbf{k}})~,
\end{equation}
where
\begin{equation}
    F_l^\mathcal{I}(k) = \left(\frac{1}{4}\mathcal{T}_{\delta_r}^\mathcal{I} + \mathcal{T}_\Psi^\mathcal{I}\right)j_l(kr) + \int_{0}^{r_\text{dec}} dr\ (\mathcal{T}_\Phi^\mathcal{I} + \mathcal{T}_\Psi^\mathcal{I})' j_l(kr) + \mathcal{T}_v^\mathcal{I} j_l'(kr) - \mathcal{T}_v^\mathcal{I}(0)j_l'(0)~.
\end{equation}
For simplicity, we consider a combination of a superhorizon adiabatic mode and a CDM isocurvature mode. Neglecting higher-order terms and evaluating the system at the time of decoupling, we use the CLASS code~\cite{Lesgourgues2011, di_dio_classgal_2013} to numerically compute the first three multipole moments as
\begin{align}
  D_1 &= -(kr_\text{dec})\cdot0.27 S_\text{dec}~, \\
  D_2 &= -(kr_\text{dec})^2\cdot(0.33 S_\text{dec}+0.32\Phi_\text{dec})~, \\
  D_3 &= -(kr_\text{dec})^3\cdot(0.33 S_\text{dec}+0.34\Phi_\text{dec})~,
\end{align}
which correspond to the dipole, quadrupole, and octopole moments, respectively.

\begin{figure}[htbp]
  \centering
  \includegraphics[width=0.49\textwidth]{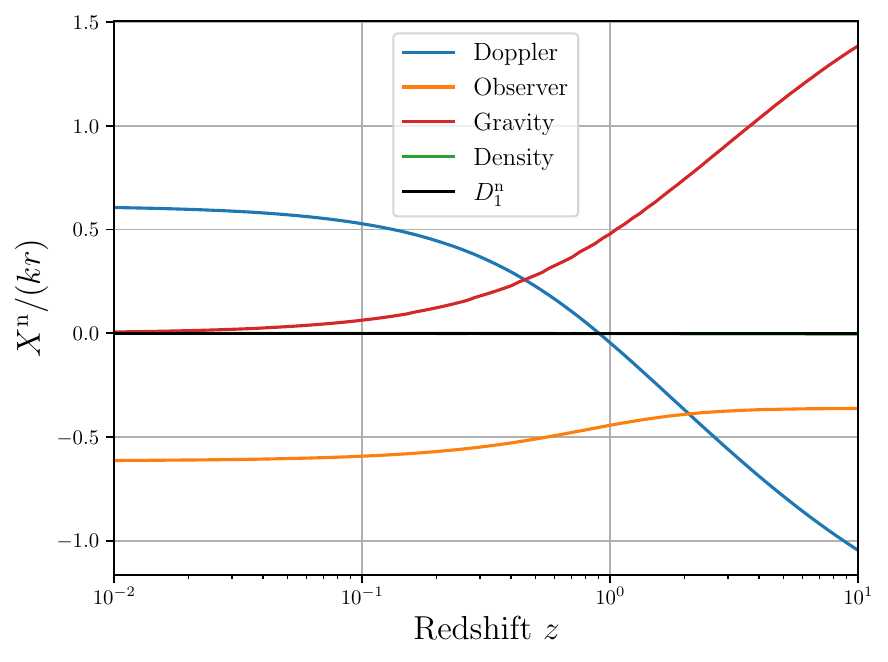}
  \includegraphics[width=0.49\textwidth]{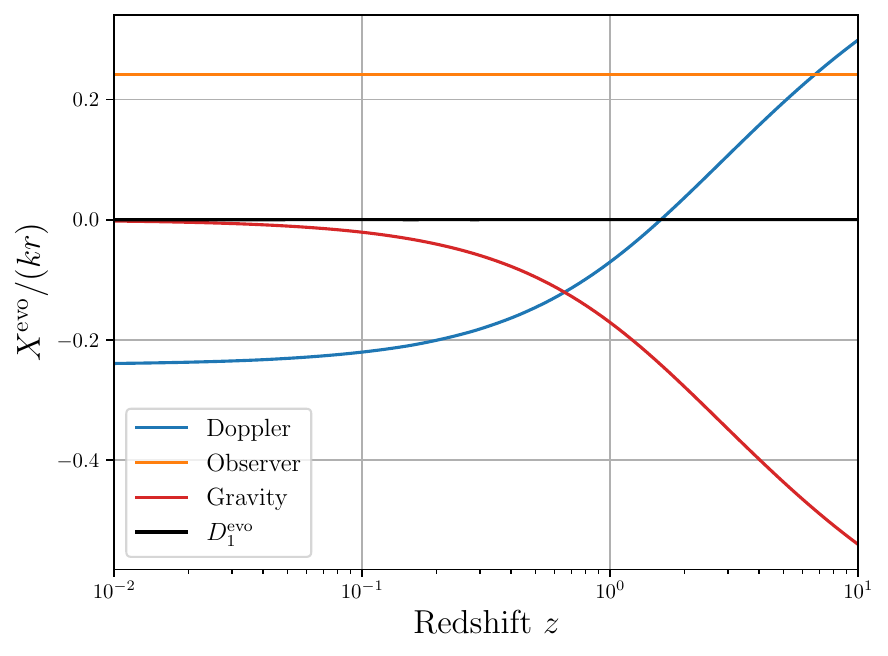} \\
  \includegraphics[width=0.49\textwidth]{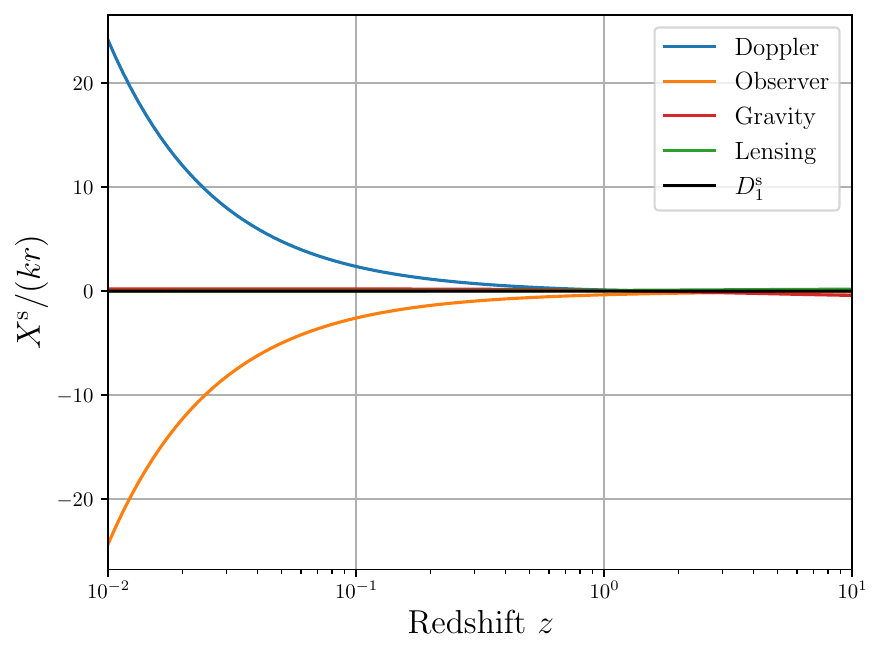}
  \includegraphics[width=0.49\textwidth]{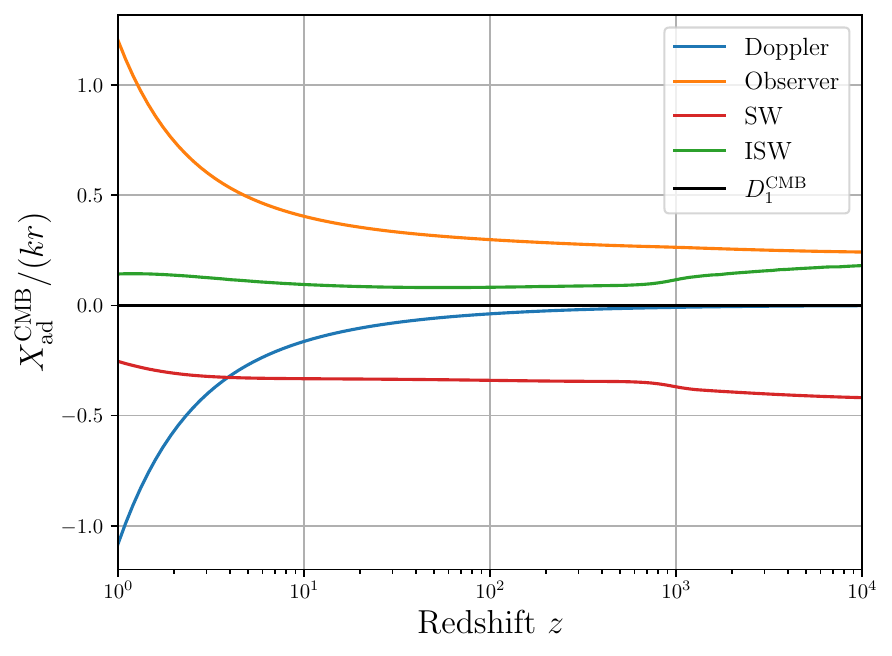} 
  \caption{\label{fig:1} Different contributions to the number count dipole and the CMB dipole of a single Fourier mode under adiabatic initial conditions. 
  The first three panels illustrate the contributions to $D_1^{n}$~[top-left], $D_1^{\text{evo}}$~[top-right] and $D_1^{s}$~[bottom-left] as defined in Eq.~(\ref{equ:1}).
  The last panel shows the contributions to $D_1^{\text{CMB}}$~[bottom-right] from the Sachs-Wolfe effect, the Doppler effect, integrated Sachs-Wolfe effect and the observer term, respectively.
  It is noteworthy that an exact cancellation occurs, resulting in $D_1 = 0$~(black) at leading order in both CMB dipole and number count dipole for adiabatic initial conditions.}
\end{figure}

As shown in the bottom-right panel of Fig.\ref{fig:1}, the leading-order intrinsic CMB dipole generated by adiabatic perturbations is canceled by the kinematic contribution, leaving only the initial isocurvature component, which can affect the CMB dipole.
The coefficients associated with the adiabatic mode are consistent with those reported in \cite{erickcek_superhorizon_2008, domenech_galaxy_2022}. However, the coefficients for the isocurvature mode are approximately 1.5 times larger than those calculated in \cite{domenech_galaxy_2022}.

\subsection{Galaxy number count fluctuations}

We now proceed to study the effect of such a monochromatic superhorizon mode on galaxy number count fluctuations.
Since the isocurvature perturbation has not fully converted into curvature perturbations by today, a residual CDM isocurvature density fluctuation still remains.
However, at redshifts $z \sim 1$, the density contrast is suppressed by a factor of $a_0/a_\text{eq} \sim 10^3$.
Therefore, the initial isocurvature perturbation can be safely neglected when evaluating its impact on galaxy number counts~\cite{domenech_galaxy_2022}.We can now focus only on the adiabatic mode for galaxy number count fluctuations, and write the spherical harmonic coefficient $a_{lm}$ as
\begin{equation}
  a_{lm}(\tau_S) = 4\pi i^l F_l^\text{ad}(\tau_S,k) \Phi_\text{ini} Y_{lm}^*(\hat{\mathbf{k}})~,
\end{equation}
where~\cite{bonvin_what_2011}
\begin{equation}
  \begin{aligned}
    F_{l}(\tau_S,k)& = j_l(kr_S)\left[b\mathcal{T}_\delta+\left(\frac{\mathcal{H}'}{\mathcal{H}^2}+\frac{2-5s}{r_S\mathcal{H}}+5s-f_\text{evo}+1\right)\mathcal{T}_\Psi + (-2+5s)\mathcal{T}_\Phi+\mathcal{H}^{-1}\mathcal{T}_\Phi^{\prime}\right] \\
    &+ \left[j_l'(kr_S)\left(\frac{\mathcal{H}'}{\mathcal{H}^2}+\frac{2-5s}{r_S\mathcal{H}}+5s-f_{\text{evo}}\right)+j_l''(kr_S)\frac{k}{\mathcal{H}} +(f_\text{evo}-3)j_l(kr_S)\frac{\mathcal{H}}{k}\right]\mathcal{T}_V \\
    &+ \frac{2-5s}{2r_S}\int_0^{r_S} dr\ j_l(kr_S)(\mathcal{T}_\Phi+\mathcal{T}_\Psi)\left[l(l+1)\frac{r_S-r}r+2\right] \\
    &+ \int_0^{r_S} dr\ j_l(kr_S)(\mathcal{T}_\Phi'+\mathcal{T}_\Psi')\left(\frac{\mathcal{H}'}{\mathcal{H}^2}+\frac{2-5s}{r_S\mathcal{H}}+5s-f_{\text{evo}}\right) \\
    &- j_l^{\prime}(0)\left(2+\frac{\mathcal{H}^{\prime}}{\mathcal{H}^2}+\frac{2-5s}{r_S\mathcal{H}}-f_{\text{evo}}\right)\mathcal{T}_V(0)~.
    \end{aligned}
    \label{equ:poles_NC}
\end{equation}
Following \cite{domenech_galaxy_2022}, we rewrite Eq.(\ref{equ:poles_NC}) in terms of three separate contributions:
\begin{equation}
F_l(\tau_S, k) = F_l^{(n)} + f_\text{evo} F_l^{(\text{evo})} + (2 - 5s) F_l^{(s)}~,
\label{equ:1}
\end{equation}
which allows us to compute each term independently of the bias $b$, $s$ and $f_\text{evo}$.  As shown in Fig.~\ref{fig:1}, under adiabatic initial conditions, a cancellation at leading order occurs within both the galaxy number count dipole and the CMB dipole.

Thus, to account for the dipole anomaly, our peculiar velocity can be inferred from the galaxy number count dipole as
\begin{align}
v_\odot^i = 797 \pm 87~ \text{km/s}~,
\end{align}
which in turn induces a kinematic CMB dipole of
\begin{align}
D_{1,\rm kin}^{\rm CMB} = (2.7 \pm 0.29) \times 10^{-3}~.
\end{align}
In comparison, the observed CMB dipole is
\begin{align}
D_1^{\rm CMB} = 1.2 \times 10^{-3}~.
\end{align}
To reconcile this difference, an intrinsic CMB dipole contribution is required:
\begin{align}
D_{1,\rm in}^{\rm CMB} \sim 1.5 \times 10^{-3}~.
\end{align}
Using the relation
\begin{align}
D_1 = - (k r_{\rm dec}) \cdot 0.27 S_{\rm dec}~,
\end{align}
we find that for an isocurvature mode with $S\sim 1$, the wavenumber must satisfy $k r_{\rm dec} \sim 5 \times 10^{-3}$ in order to explain the dipole anomaly.

\section{\label{sec:continuous spectrum}THE EFFECT OF CONTINUOUS SPECTRUM}
In realistic models, perturbations are more likely to exhibit a continuous power spectrum rather than being dominated by a single mode. 
For a continuous spectrum, to compute the spherical harmonic coefficients, $a_{lm}$, we need to integrate over all possible wavevectors, $\mathbf{k}$. 
This yields
\begin{equation}
    \begin{aligned}
      a_{lm}^\mathcal{I} &= 4\pi i^l \int\frac{d^3\mathbf{k}}{(2\pi)^3}X^\mathcal{I}(\mathbf{k}) \mathcal{T}_l^\mathcal{I}(k)Y_{lm}^*(\hat{\mathbf{k}})~, \\
      a_{lm} &= 4\pi i^l \int\frac{d^3\mathbf{k}}{(2\pi)^3}\left(\sum_\mathcal{I} X^\mathcal{I}(\mathbf{k}) \mathcal{T}_l^\mathcal{I}(k)\right)Y_{lm}^*(\hat{\mathbf{k}})~.
    \end{aligned}
\end{equation}
The power spectrum of variable $X$ between modes with initial condition $\mathcal{I}$ and $\mathcal{J}$ can be define as
\begin{equation}
  k^3\braket{X^\mathcal{I}(\mathbf{k})X^\mathcal{J}(\mathbf{k}')} = (2\pi)^3 \delta^{(3)}(\mathbf{k}-\mathbf{k}') \mathcal{P}_{\mathcal{I}\mathcal{J}}(k)~,
\end{equation}
A straightforward calculation then yields the angular power spectra as
\begin{equation}
    \begin{aligned}
      C_l^\mathcal{IJ} &= \frac{2}{\pi} \int_0^\infty \frac{dk}{k} \mathcal{P}_\mathcal{IJ}(k) F_l^\mathcal{I}(k) F_l^\mathcal{J}(k), \\
      C_l &= \frac{2}{\pi} \int_0^\infty \frac{dk}{k} \left(\sum_\mathcal{I,J} \mathcal{P}_\mathcal{IJ}(k)\right) F_l^\mathcal{I}(k) F_l^\mathcal{J}(k)~.
    \end{aligned}
\end{equation}
where $\mathcal{T}^\mathcal{I}$ and $\mathcal{T}^\mathcal{J}$ represent the transfer functions corresponding to the modes with initial conditions $\mathcal{I}$ and $\mathcal{J}$, respectively.

For simplicity, we focus solely on the isocurvature mode and define the power spectrum of the superhorizon isocurvature perturbation as
\begin{equation}
k^3 \braket{S(\mathbf{k})S(\mathbf{k}')} = (2\pi)^3 \delta^{(3)}(\mathbf{k}-\mathbf{k}')\mathcal{P}_S(k)~.
\end{equation}
In this paper, we consider two types of isocurvature power spectra: one with a power-law form and the other motivated by axion-induced perturbations.

\subsection{Constraints from CMB}
The current measurements of CMB impose stringent constraints on the amplitude of uncorrelated dark matter isocurvature perturbations, which can be quantified as follows:
\begin{equation}
  \beta \equiv \frac{\mathcal{P}_S(k)}{\mathcal{P}_R(k)}\bigg|_{k=k_\mathrm{pivot}} < 0.038~,
\end{equation}
where $\mathcal{P}_R(k)$ denotes the primordial power spectrum and $k_\mathrm{pivot}$ is the pivot scale.
The Planck data imposes a constraint on the amplitude of dark matter isocurvature perturbations, requiring that the power spectrum of such perturbations satisfies $P_S \lesssim 0.8 \times 10^{-10}$ at the wavenumber $k_\mathrm{low} = 0.002\ \text{Mpc}^{-1}$. This is known as the isocurvature limit, indicated by the corresponding red exclusion region in Fig.~\ref{fig:3}.

Multipole measurements impose important additional constraints on the power spectrum of isocurvature perturbations. In particular, the quadrupole moment provides the most stringent bounds for superhorizon perturbations, whereas constraints from higher-order multipoles are considerably weaker~\cite{han_qcd_2023}. Notably, the Planck measurements~\cite{planck_collaboration_planck_2020, collaboration_planck_2021} report a quadrupole value of $\mathcal D_2 = 2.26^{+5.33}_{-1.32} \times 10^2\ \mu\text{K}^2$, while the best-fit $\Lambda$CDM model predicts $ \mathcal  D_2 = 1016\ \mu\text{K}^2$~\cite{collaboration_planck_2021}.  Although the discrepancy indicates that the observed quadrupole is significantly lower than the standard model expectation, these measurements still provide meaningful constraints on the quadrupole within the context of our model.

When the quadrupole arises from two different power spectra, their combination must be taken into account. Let $a_{2m}^{(1)}$ and $a_{2m}^{(2)}$ denote the spherical harmonic coefficients corresponding to the two power spectra, $P_1(k)$ and $P_2(k)$, respectively. The combined quadrupole coefficients are given by
\begin{equation}
a_{2m} = a_{2m}^{(1)} + a_{2m}^{(2)}~.
\end{equation}
Accordingly, the combined quadrupole power spectrum $D_2$ is
\begin{equation}
 \mathcal  D_2 =\mathcal  D_2^{(1)} +\mathcal  D_2^{(2)}~.
\end{equation}

Give the observation $\mathcal  D_2^{\rm obs}\simeq 226 \mu \mathrm{K}^2$, we derive the 2$\sigma$ upper limits on the quadrupole amplitude as follows:
\begin{equation}
\mathcal  D_2 \lesssim 1.4 \times 10^{3} {\mu \mathrm{K}^2}~.
\end{equation}
Note the $D_2$ defined here is related to the $C_2$ in our calculation as follows,
\begin{equation}
\mathcal  D_2 = \frac{2(2+1)C_2}{2\pi T_0^2}~.
\end{equation}
where $T_0\sim 2.7$ K is the average temperature of CMB. This upper limit translates into an upper bound on the quadrupole component $C_2^\text{iso}$ generated by the superhorizon isocurvature mode,
\begin{equation}
C_2^\text{iso} < 5.5\times 10^{-11}~,
\end{equation}
which is corresponded to the blue exclusion region in Fig.~\ref{fig:3}.

\subsection{The power-law spectrum}
For the first case, $\mathcal{P}_S(k)$ takes the form of a power-law spectrum, expressed as
\begin{equation}
   \mathcal{P}_S(k) = A_I (k/k_\text{max})^{n-1}\theta(k_\text{max}-k)~,
\end{equation}
where $A_I$ is the amplitude, $k_\text{max}$ is the maximum comoving wave number of isocurvature perturbations, $n$ is the spectral index which is set to be 1 in our calculation and $\theta(k)$ is the step function. Therefore the angular power spectrum $C_l$ can be written as
\begin{equation}
  C_l = \frac{2}{\pi} \int_0^{k_\text{max}} \frac{dk}{k} \mathcal{P}_S(k) |F_l(k)|^2~.
\end{equation}

\subsection{The axion spectrum}
\begin{figure}[htbp]
  \centering
  \includegraphics[width=0.485\textwidth]{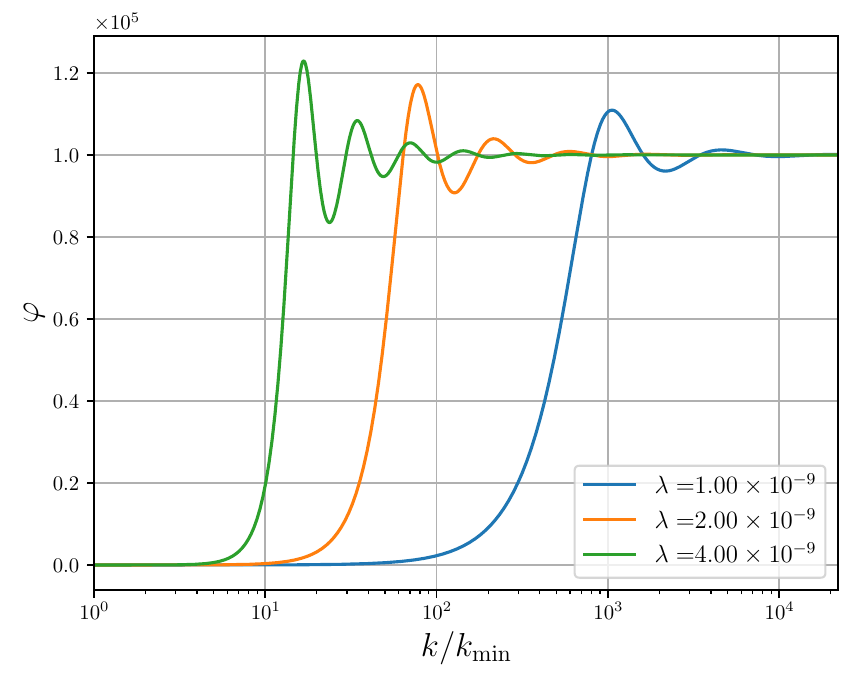}
  \includegraphics[width=0.5\textwidth]{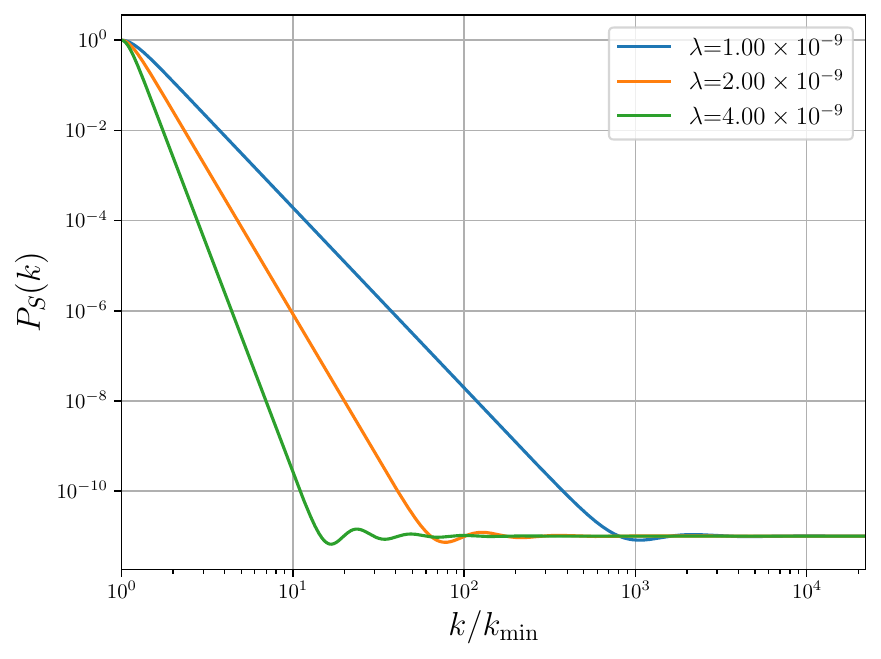}
  \caption{\label{fig:2} [Left]~Evolution of the radial mode $\varphi$ during inflation for $f=1\times10^5$ and $\lambda = 1 \times 10^{-9}, 2 \times 10^{-9},$ and $4 \times 10^{-9}$.
  [Right]~Power spectrum $\mathcal{P}_S(k)$ of the isocurvature perturbation induced by axion for $f=10^5 H$ and $\lambda = 1 \times 10^{-9}, 2 \times 10^{-9},$ and $4 \times 10^{-9}$.}
\end{figure}

The second power spectrum is given by~\cite{han_qcd_2023}
\begin{equation}
  \mathcal{P}_S(k) = \mathcal{P}_S(k_\text{min}) \left[\frac{\varphi(k)}{\varphi(k_\text{min})}\right]^{-2}\theta(k-k_\text{min})~,
\end{equation}
where $\mathcal{P}_S(k_\text{min}) \approx 1$, $k_\text{min}$ is the minimum comoving wave number at which the isocurvature perturbation becomes relevant, and $\varphi(k)$ is the radial mode of axion governed by
\begin{equation}
  \ddot{\varphi} + 3H\dot{\varphi} + V'(\varphi) = 0, \quad V(\varphi) = \frac{\lambda}{4}\left(\varphi^2 - f^2\right)^2~,
\end{equation}
where $f$ is the decay constant of the axion and $\varphi$ is initially displaced slightly from the origin, with a characteristic scale of displacement around $H$. If this displacement occurs at the superhorizon scale $k_\text{min}$, the resulting isocurvature perturbation is enhanced at the superhorizon scale while remaining small at the scale $k_\text{dec}$.
For this case, the CMB power spectrum can be written as
\begin{equation}
  C_l = \frac{2}{\pi} \int_{k_\text{min}} \frac{dk}{k} \mathcal{P}_S(k) |F_l(k)|^2~.
\end{equation}
In the left panel of Fig.~\ref{fig:2}, we show the evolution of $\varphi$ for different values of the parameter $\lambda$. Considering the constraints imposed by current CMB observations~\cite{collaboration_planck_2021}, which lead to the bound $H/f \lesssim 10^{-5}$ for $\theta_0 \sim O(1)$, the initial conditions are specified as $\varphi = 0$, $\varphi(k_\text{min}) = H/\pi$, and $f = 10^5H$. The right panel presents the corresponding axion isocurvature power spectrum $\mathcal{P}_S(k)$. It is observed that as the field attains the value of $f$, the power spectrum $\mathcal{P}_S(k)$ diminishes and almost vanishes. When $\lambda$ is too small, the power spectrum tends to maintain relatively high values over the extended duration, leading to significant isocurvature perturbations at the epoch of recombination, which breaks the isocurvature limit. On the other hand, if $\lambda$ is excessively large, the field rapidly reaches the value of $f$, potentially resulting in a dipole generated by the power spectrum that is insufficient to solve the cosmic dipole problem.

\subsection{Numerical result}

\begin{figure}[htbp]
  \centering
  \includegraphics[width=0.49\textwidth]{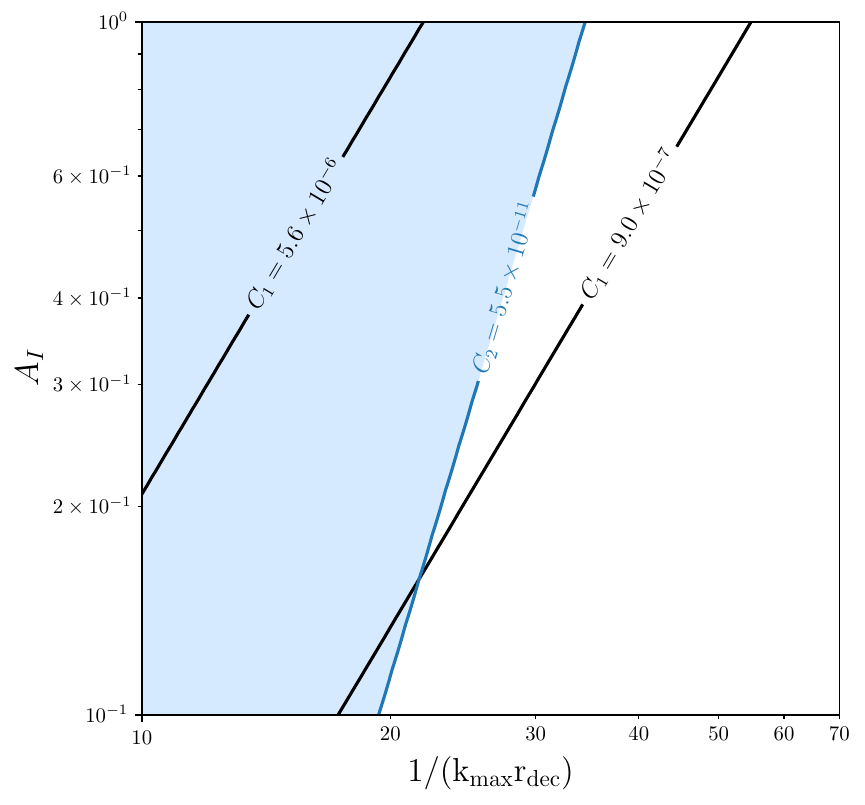}
  \includegraphics[width=0.49\textwidth]{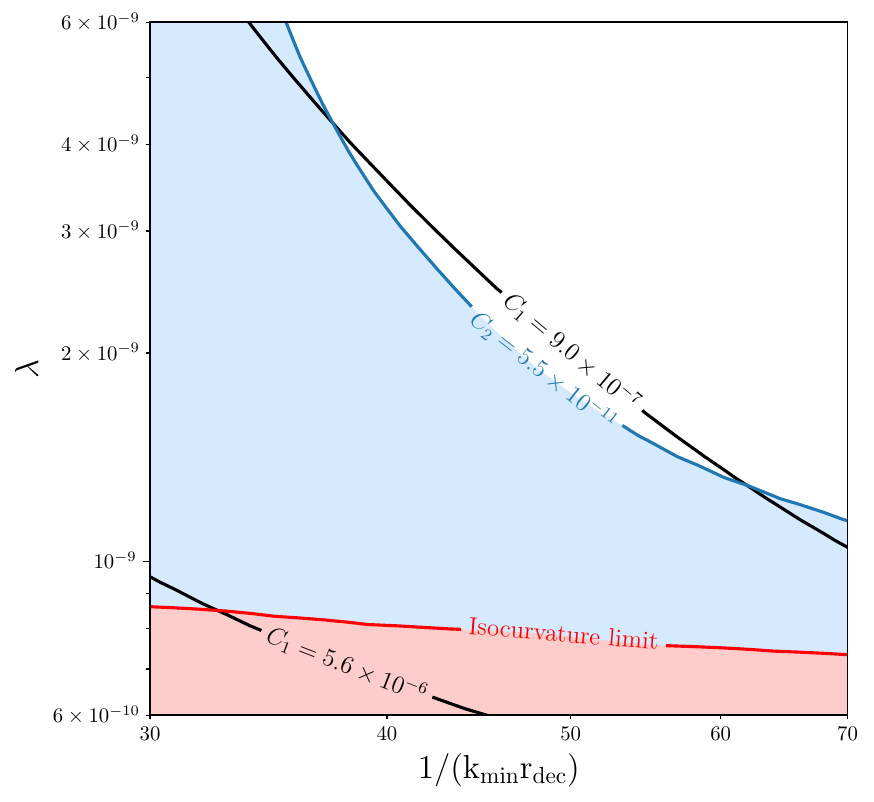}
  \caption{\label{fig:3} Parameter spaces of the two models to explain the comic dipole problem. The two black curves represent the lower and upper limits of the dipole, respectively. The blue regions indicate the $2\sigma$ exclusion limits from the observed quadrupole $C_2$. While the red region is excluded, as it would result in excessively large isocurvature perturbations.
  [Left]~Parameter space of the power-law spectrum.
  [Right]~Parameter space of the axion power spectrum.}
\end{figure}

We numerically compute the angular power spectra of the CMB and number count fluctuations. Similar to the single mode case, the number count dipole and the CMB dipole vanish for adiabatic initial conditions, while the isocurvature CMB dipole does not. To address the dipole tension, we plot the parameter space for both scenarios in Fig.~\ref{fig:3}, where the two black curves represent the lower and upper bounds on the dipole amplitude, and the blue regions indicate the $2\sigma$ exclusion limits from the observed quadrupole $C_2$. 

As shown in the left panel of Fig.~\ref{fig:3}, we present the parameter space required to explain the cosmic dipole anomaly in the $1/(k_\mathrm{max}r_\mathrm{dec})$–$A_I$ plane. In this analysis, the isocurvature constraint is imposed by restricting the parameter space to the region where $k_\mathrm{max} < k_\mathrm{dec}$. By focusing on this region, we avoid the need to explicitly display the isocurvature bound, as it is automatically satisfied within the selected parameter space. We find that explaining the dipole anomaly requires $A_I \gtrsim 0.2$; otherwise, the predicted CMB dipole is too small or the quadrupole becomes too large.

In the right panel of Fig.~\ref{fig:3}, we present the parameter space required to explain the cosmic dipole tension in the $1/(k_{\min}r_\mathrm{dec})$–$\lambda$ plane. The red region is excluded due to large isocurvature perturbations. We find that the self-coupling must lie within the range $10^{-9} < \lambda < 4 \times 10^{-9}$. If the axion decay constant $f$ is determined by future observations of axion couplings, the corresponding scalar mass associated with the axion can also be inferred. For example, if $f \sim 10^{11}$ GeV, the predicted scalar mass is expected to be around $\mathcal{O}(10^6)$ GeV. Although this mass is beyond the reach of the current LHC, a future 1000 TeV collider may be capable of detecting this particle.

\section{CONCLUSIONS}
In this work, we have investigated both discrete and continuous power spectra of superhorizon cold dark matter (CDM) isocurvature perturbations as a potential resolution to the cosmic dipole tension. Our analysis of a single plane-wave perturbation shows that an initial adiabatic superhorizon mode does not contribute to either the CMB dipole or the galaxy number count dipole. This is due to the fact that such adiabatic perturbations cancel at leading order, and superhorizon isocurvature perturbations are converted into adiabatic modes during the matter-dominated era, rendering them ineffective in altering the number count dipole. However, a superhorizon isocurvature mode can induce a negative intrinsic dipole in the CMB, partially canceling the kinematic dipole from our peculiar motion and thereby helping to reconcile the observed discrepancy.

We then extended our analysis to continuous spectra of isocurvature fluctuations. Both nearly scale-invariant power-law spectra and axion-generated spectra were found to be capable of resolving the dipole anomaly while remaining consistent with current CMB constraints. For the power-law case, we find that explaining the observed dipole requires $A_I \gtrsim 0.2$ for a suitable cutoff scale $k_\text{max}$. In the axion scenario, the evolution of the radial mode during inflation naturally generates large superhorizon isocurvature perturbations that decay at recombination scales. We find that for $f = 10^5H$ and $10^{-9} < \lambda < 4 \times 10^{-9}$ with an appropriate choice of $k_\text{min}$, the observed dipole tension can be effectively resolved.

If confirmed, this mechanism would not only address the cosmic dipole problem but also offer indirect support for the existence of axion dark matter. Our results thus preserve the cosmological principle while motivating further investigation into superhorizon fluctuation generation and the broader implications of axion physics in the early Universe.

\vspace*{3mm}
\noindent 
\textbf{Acknowledgments}\\[0.5mm] 
 The work is supported by the National Key R{\&}D Program of China under grant 2023YFA1606100 and by the National Natural Science Foundation 
of China under grants No. 12435005. 
C.\,H.\ acknowledges supports from the Sun Yat-Sen University Science Foundation, 
the Fundamental Research Funds for the Central Universities at Sun Yat-sen University under Grant No.\,24qnpy117, and the Key Laboratory of Particle Astrophysics and Cosmology (MOE) of Shanghai Jiao Tong University.\

\addcontentsline{toc}{section}{References}
\bibliographystyle{JHEP}
\bibliography{references}



\end{document}